\documentclass[onecolumn,prl,twocolumn,superscriptaddress]{revtex4}
\usepackage[latin9]{inputenc}
\usepackage{color}
\usepackage{amsmath}
\usepackage{amssymb}
\usepackage{graphicx}
\usepackage{tikz}
\usepackage{amsfonts}
\usepackage{graphicx} 
\usepackage{dsfont}
\usepackage{algorithm}
\usepackage{algorithmic}
\usepackage{multirow}

\usepackage{mathrsfs} 
\usepackage{float}
\usepackage{bbm}
\usepackage{comment}

\usepackage{epstopdf}
\usepackage{epsfig} 
\usepackage{verbatim}
\usepackage{array}
\usepackage{setspace}

   \allowdisplaybreaks[1]
   
\usepackage[unicode=true,
 bookmarks=true,bookmarksnumbered=false,bookmarksopen=false,
 breaklinks=false,pdfborder={0 0 1},backref=false,colorlinks=true]
 {hyperref}
\hypersetup{
 linkcolor=magenta, urlcolor=blue, citecolor=blue, pdfstartview={FitH}, hyperfootnotes=false, unicode=true}

\makeatletter
\@ifundefined{textcolor}{}
{%
 \definecolor{BLACK}{gray}{0}
 \definecolor{WHITE}{gray}{1}
 \definecolor{RED}{rgb}{1,0,0}
 \definecolor{GREEN}{rgb}{0,1,0}
 \definecolor{BLUE}{rgb}{0,0,1}
 \definecolor{CYAN}{cmyk}{1,0,0,0}
 \definecolor{MAGENTA}{cmyk}{0,1,0,0}
 \definecolor{YELLOW}{cmyk}{0,0,1,0}
}

\usepackage{amsfonts}\usepackage{tabularx}\usepackage{dcolumn}\usepackage{bm}\usepackage{graphicx}\usepackage{epstopdf}

\hypersetup{urlcolor=blue}

\makeatother

\newcolumntype{C}[1]{>{\centering\arraybackslash$}p{#1}<{$}}

\begin{document}

\title{Optimizing Quantum Transformation Matrices: A Block Decomposition Approach for Efficient Gate Reduction}

\author{Kin Man Lai}
\affiliation{Department of Physics, City University of Hong Kong, Tat Chee Avenue, Kowloon, Hong Kong SAR, China}
\author{Xin Wang}
\email{x.wang@cityu.edu.hk}
\affiliation{Department of Physics, City University of Hong Kong, Tat Chee Avenue, Kowloon, Hong Kong SAR, China}
\affiliation{City University of Hong Kong Shenzhen Research Institute, Shenzhen, Guangdong 518057, China}
\affiliation{Quantum Science Center of Guangdong-Hong Kong-Macao Greater Bay Area, Shenzhen, Guangdong 518045, China}

\date{\today}

\begin{abstract}
This paper introduces an algorithm designed to approximate quantum transformation matrix with a restricted number of gates by using the block decomposition technique. Addressing challenges posed by numerous gates in handling large qubit transformations, the algorithm provides a solution by optimizing gate usage while maintaining computational accuracy. Inspired by the Block Decompose algorithm, our approach processes transformation matrices in a block-wise manner, enabling users to specify the desired gate count for flexibility in resource allocation. Simulations validate the effectiveness of the algorithm in approximating transformations with significantly fewer gates, enhancing quantum computing efficiency for complex calculations.
\end{abstract}

\maketitle

\section{Introduction}
Quantum information processing relies on the precise and efficient manipulation of quantum states. Quantum state transformation aims to identify a circuit that can guide one quantum state to another. In particular, when provided with an initial state $| \psi_i \rangle$ and a target state $| \psi_t\rangle$, the objective of quantum state transformation is to find an optimal matrix transformation $\mathbf{U}$ that minimizes the difference between $\mathbf{U}| \psi_i\rangle$ and $|\psi_t\rangle$.

One significant challenge in quantum state transformation is the exponential increase in the quantum state dimension as the number of qubits rises, necessitating a greater number of gates to construct the required transformation matrix. Ref.~\cite{nielsen2010quantum} provides insights into the complexities associated with quantum computations and the exponential scaling of quantum state spaces. Additionally, Ref.~\cite{man2024group} highlights the enormous number of gates needed for transformations involving large qubits even with the introduction of sparsity, underscoring the need for more efficient approaches.

Decomposition techniques play a crucial role in simplifying quantum circuits by breaking down extensive transformation matrices into more manageable sequences of gates. By deconstructing a transformation matrix representing a specific operation into a quantum circuit, these methods aim to disassemble unitary transformation matrices into sequences of elementary gates. Numerous papers, such as~\cite{Vartiainen_2004, mottonen2004quantum, shende2005synthesis, mottonen2006decompositions, li2013decomposition, li2014decomposition, di2013synthesis, geller2021experimental, malvetti2021quantum} focus on decomposing the unitary transformation matrix into elementary gates, effectively simplifying circuit complexity. While these methods have shown effectiveness, they can sometimes introduce additional complexity in terms of circuit depth, necessitating further optimization processes.

In light of these challenges, gate reduction emerges as a vital process for efficient quantum manipulation. Gate reduction within quantum circuits aims to minimize the number of quantum gates required to perform a given computation. Recent papers have demonstrated success in gate reduction strategies. For instance, Ref.~\cite{abdessaied2014quantum} explores the use of quantum gate libraries to optimize and reduce the gate count in quantum circuits. 
Ref.~\cite{Nam_2018} discusses an automated optimization approach that continuously refines large quantum circuits to reduce gate counts. Ref.~\cite{duncan2020graph, kissinger2020reducing} successfully reduced gate counts by incorporating ZX-calculus techniques for circuit simplification. Recently, Ref.~\cite{Plesch2011, zhang2023, zhang2024} have developed algorithms aimed at reducing the depth of quantum circuits representing diagonal unitary matrices by minimizing the number of Z-rotation gates and CNOT gates required.

The block decomposition algorithm introduced by Ref.~\cite{yuan2020block} has demonstrated remarkable success in sparse optimization. Due to its unique combinatorial approach, this method surpasses other optimization methods, such as alternating direction methods~\cite{song2016alternating, xu2016empirical} or decomposition methods~\cite{hamidi2010fast, zarabie2019l0normconstrainednonnegativematrix, lu2010penalty} on similar problems. Building upon the strengths of this method, our objective is to improve the efficiency of quantum circuits by carefully choosing the most precise gates to represent the circuit through exhaustive optimization for gate selection. By breaking down the transformation into manageable blocks, we aim to significantly reduce the number of gates required to construct given transformation matrices.

In this paper, we focus on developing an algorithm that approximates a given transformation operator as a limited number of gates using the block decomposition method. The key objective is to bridge the gap between the success of the block decomposition algorithm in sparse optimization, providing a new approach to addressing the challenges with large qubit manipulation. By integrating these two areas, we aim to introduce flexibility in gate construction while maintaining the ability to limit the gate count.

The remainder of this paper is organized as follows: 
We begin by presenting the mathematical formulations of the problem in Section~\ref{sec:formulation}. Our block decomposition method for searching a limited number of gates to model the target transformation matrix is presented in Section~\ref{sec:method}. In Section~\ref{sec:result}, we present our comparative analysis and numerical results. Section~\ref{sec:discussion} delves into the significance of our work. Finally, we conclude the paper in Section~\ref{sec:conclusion} with a summary of our work and future research directions.


\section{Problem Formulation} \label{sec:formulation}

Consider a given transformation matrix $\mathbf{U}$ for a certain quantum operation. This matrix can be decomposed into a series of $n$ gates forming a quantum circuit, i.e., $\mathbf{U} = \mathbf{U}_1^\dag \mathbf{U}_2^\dag \mathbf{U}_3^\dag \dots \mathbf{U}_n^\dag$, using the cosine-sine decomposition method~\cite{mottonen2004quantum, di2013synthesis}. Furthermore, each of these $2^n \times 2^n$ decomposed matrices, $\mathbf{U}_k^\dag$, can be converted into a $2\times 2$ sub-matrix, $\mathbf{u}_k$. By examining each $\mathbf{u}_k$ gate individually, it can be represented using a sequence of $R(z) R(y) R(z)$ gates:

\begin{equation} \label{eq:rotation}
\mathbf{u}_k = \begin{bmatrix}
\alpha & \beta \\
\gamma & \delta 
\end{bmatrix}
= R_z(\theta) \cdot R_y(\phi) \cdot R_z(\lambda) \;\; ,
\end{equation}
where $\theta$, $\phi$, and $\lambda$ are the corresponding Euler angles. {\color{black}Here, the rotation gates $R_z(\theta)$ and $R_y(\phi)$ are defined as:

\begin{equation} \label{eq:rotation_Rzy}
R_z(\theta) = \begin{bmatrix}
e^{-i\theta/2} & 0 \\
0 & e^{i\theta/2}
\end{bmatrix} \;\;\;\;\; \text{and} \quad
R_y(\phi) = \begin{bmatrix}
\cos(\phi/2) & -\sin(\phi/2) \\
\sin(\phi/2) & \cos(\phi/2)
\end{bmatrix},
\end{equation}
which ensures that each gate is unitary with a determinant of 1. Consequently, the overall unitary matrix $\mathbf{U}$ is also restricted to have a determinant of 1, preserving the unitary property of the transformation. }Appendix~\ref{appx:appxA} provides insights on the positioning of these unitary matrices within the transformation matrix and their relationship with the qubits targeted by the gates.

To minimize the number of gates required to construct a given matrix $\mathbf{U}$, we aim to find a new transformation matrix $\mathbf{Y}$ that approximates $\mathbf{U}$ using a limited number of $\mathbf{U}_i^\dag$. To achieve this, we seek to decrease the discrepancy between the two transformation matrices using an error function,
\begin{equation} \label{eq:error_func}
\mathrm{arg \min_{\mathbf{Y}}}\;\; \frac{1}{2}\left\lVert\mathbf{Y}-\mathbf{U}\right\rVert^2_2 \;\; .
\end{equation}

Note that $ \|\mathbf{Y}-\mathbf{U}\|_2^2$ is defined as the square of the Euclidean norm of $\mathbf{Y}-\mathbf{U}$, where the $\frac{1}{2}$ factor is added for convenience in calculations~\cite{man2024group}.  This expression commonly serves as a fidelity term in optimization contexts.  Note that Euclidean norm is given by the following equation:
\begin{equation} \label{eq:l2norm}
  \|\mathbf{X}\|_2 = \left(\sum_{i=1}^m\sum_{j=1}^n |x_{ij}|^2 \right)^{\frac{1}{2}} \;\;.
\end{equation}

By optimizing Eq.~\eqref{eq:error_func}, we obtain an optimized $\mathbf{Y}$ that can approximate the transformation operator, $\mathbf{U}$, formed by a series of $\mathbf{Y}_k^\dag$.

For simplicity, we define $\mathbf{X}_k = \mathbf{Y}_k^\dag$, such that $\mathbf{Y} = \mathbf{X}_1 \mathbf{X}_2 \mathbf{X}_3 \dots \mathbf{X}_M = \prod_{k=1}^{M} \mathbf{X}_k$. Each $\mathbf{X}_k$ represents a $2^n \times 2^n$ decomposed matrix from $\mathbf{Y}$, suitable for representation as a quantum gate in a quantum circuit, which can each be converted into the corresponding $2\times 2$ sub-matrix,$\mathbf{x}_k$. We term these $\mathbf{X}_k$ matrices as gate matrices and restrict the number of $\mathbf{X}_k$ to a maximum of $M$. This limitation on $\mathbf{X}$ allows the transformation matrix $\mathbf{Y}$ to be constructed with a desired number of gates essential for a specific transformation, ensuring manageable complexity in quantum computation. By expanding $\mathbf{Y}$, we can rewrite Eq.~\eqref{eq:error_func} as:

\begin{equation} \label{eq:error_func2}
\mathrm{arg \min_{\prod_{k=1}^{M} \mathbf{X}_k}} \;\; \frac{1}{2}\left\lVert\mathbf{X}_1 \mathbf{X}_2 \mathbf{X}_3 \dots  \mathbf{X}_M -\mathbf{U}\right\rVert^2_2 \;\;.
\end{equation}

To tackle this problem, we focus on updating only one of these sub-matrices $\mathbf{X}_k$ in each iteration. Lets designate $\mathbf{X}_w$ as the gate matrix for this iteration to be updated. We introduce two storage variables, $\mathbf{A} = \mathbf{X}_1 \mathbf{X}_2 \dots  \mathbf{X}_{w-1}$ and $\mathbf{B} = \mathbf{X}_{w+1} \mathbf{X}_{w+2} \dots  \mathbf{X}_M$. We solve Eq.~\eqref{eq:error_func2} by choosing different $w$ in each iteration and iteratively solving the following optimization problem:

\begin{equation} \label{eq:objective}
\mathrm{arg \min_{\mathbf{X}_w}} \;\; \frac{1}{2}\left\lVert\mathbf{A} \mathbf{X}_w \mathbf{B} - \mathbf{U}\right\rVert^2_2 \;\; .
\end{equation}

Furthermore, we need to consider two important properties of these transformation matrices by introducing suitable constraints. The first property is the unitary constraint, which ensures that the transformation matrix is reversible and that the probabilities sum to one. Secondly, each $X_i$ has a specific structure of being a combination of an $2 \times 2$ sub-matrix and identity matrix. Thus only four elements (i.e. $\alpha$, $\beta$, $\gamma$, $\delta$) in specific positions, namely $(i,i)$, $(i,j)$, $(j,i)$ and $(j,j)$ are distinct from being an identity matrix. 

To this end, we present our objective problem in each iteration as the following:
\begin{equation} \label{eq:main}
\mathrm{arg \min_{\mathbf{X}_w}} \;\; \frac{1}{2}\left\lVert\mathbf{A} \mathbf{X}_w \mathbf{B} - \mathbf{U}\right\rVert^2_2 \;\; \mathrm{s.t.} \;\; \mathbf{X}_w^T \mathbf{X}_w = \mathbf{I} \;\; , \;\; \mathbf{D} \mathbf{X}_w = \mathbf{D} \mathbf{I} \;\;.
\end{equation}

The first constraint, $\mathbf{X}_w^T \mathbf{X}_w = \mathbf{I}$, ensures that the updated $\mathbf{X}_w$ is a unitary transformation matrix. Since the multiplication of unitary matrices results in a unitary matrix, this guarantees that the state transformation matrix, $\mathbf{Y}$, is also a unitary matrix.

By choosing a suitable dictionary matrix $\mathbf{D}$, the second constraint, $\mathbf{D} \mathbf{X}_w = \mathbf{D} \mathbf{I}$, ensures that only four elements in specific locations within $\mathbf{X}_w$ do not correspond to elements within an identity matrix. By changing the dictionary $\mathbf{D}$, we can determine the position of the sub-matrix within $\mathbf{X}_w$. Introducing this constraint updates $\mathbf{X}_w$ through optimizing the positions and values of $\alpha$, $\beta$, $\gamma$, and $\delta$.

In the context of gate-based quantum computing, the repeated use of certain gates could lead to inefficiencies. To address this issue, particularly in the construction of the dictionary matrix $\mathbf{D}$, we have implemented a strategy that only considers gates that have not been used by the other $X_{(M \neq W)}$ gates in the circuit. By eliminating the positions in $\mathbf{D}$ that are already used by the other $M-1$ gates $X_{(M \neq W)}$, we ensure that the selected gates are not repeated. This adjustment significantly reduces the complexity required for our algorithm. Given that quadratic programming is the most computationally expensive step of this algorithm, similar to \cite{yuan2020block}, we can now proceed with $M-1$ fewer quadratic programming steps instead of considering all gates, after discarding some gates for consideration.

\section{Block decomposition method on gate minimizing} \label{sec:method}

In this section, we outline our approach to optimizing the number of gates required to construct the transformation matrix $\mathbf{U}$ by solving the objective problem defined in Eq.~\eqref{eq:main}. As Eq.~\eqref{eq:main} poses a non-convex optimization challenge due to the unitary constraint, we employ a penalty method to  address this constraint. This reformulation makes the objective more tractable by incorporating the non-convex constraint into the objective function. Through the introduction of a penalty parameter $\lambda$ to regulate the unitary constraint, the objective problem is reformulated at each iteration as follows:
\begin{equation} \label{eq:lagrangian}
\mathrm{arg \min_{\mathbf{X}_w}} \;\; \frac{1}{2}\left\lVert\mathbf{A} \mathbf{X}_w \mathbf{B} - \mathbf{U}\right\rVert^2_2 + \lambda (\mathbf{X}_w^T \mathbf{X}_w - \mathbf{I}) \;\; \mathrm{s.t.} \;\; \mathbf{D} \mathbf{X}_w = \mathbf{D} \mathbf{I} \;\;.
\end{equation}

We propose to tackle the problem using the block decomposition algorithm solve Eq.~\eqref{eq:lagrangian}, a comprehensive outline for the algorithm is provided as follows:

\begin{algorithm}
\caption{\bf Block Decomposition for Transformation Matrix Approximation}
\label{BD_algorithm}
\begin{algorithmic}[1]
  \STATE Specify $K$, the number of gate matrices $\mathbf{X}$ wanted to use to approximate $\mathbf{U}$. Define a initial $\mathbf{Y}^0$ and set $t=0$. 
  \STATE Find an initial feasible solution by decomposing initial $\mathbf{Y}^0$ as $\mathbf{X}^0_1 \mathbf{X}^0_2 \mathbf{X}^0_3 \dots  \mathbf{X}^0_K$
 
  \WHILE{not converge}
  \STATE Use some strategy to choose a gate matrix for updating $\mathbf{X_w}$ and  calculate for $\mathbf{A}$ and $\mathbf{B}$.
  \STATE Update $\mathbf{X}_w$ by solving sub-problem Eq.~\eqref{eq:lagrangian} using combinatorial search.
    \STATE Adaptively update penalty parameters $\lambda$ and $\mu$. 
  \STATE Increment $t$ by 1.
  \ENDWHILE
\end{algorithmic}
\end{algorithm}

 \subsection{Choice of $\mathbf{X}_w$ }
 
The choice of $\mathbf{X}_w$ in each iteration is arbitrary and can be approached through various methods, with common ones being the cyclic and random methods. In the cyclic method, the choice of the working gate matrix follows a predetermined order based on the lineup, offering a straightforward approach. In contrast, the random method involves selecting the gate matrix to update at random. This approach can help avoid some of the pitfalls of the cyclic method, such as getting stuck in sub-optimal cycles. Furthermore, random selection can potentially lead to faster convergence on average.

 Since $\mathbf{X}_w$ have a definite form, after choosing the working gate matrix for each iteration, we exhaustively search for optimal $\mathbf{X}_w$ that can obtain minimum objective value of Eq.~\eqref{eq:lagrangian}.
 Note that after the selection of $\mathbf{X}_w$, all other gate matrices $\mathbf{X}_{\neq w}$ are fixed for the update, considered as the constant $\mathbf{A}$ and $\mathbf{B}$ respectively in Eq.~\eqref{eq:lagrangian}.

 \subsection{Updating $\mathbf{X}_w$}
The key of the proposed algorithm lies in minimizing the objective function, Eq.~\eqref{eq:lagrangian}, for the gate matrix $\mathbf{X}_w$ in each iteration. This is achieved through an exhaustive search for the optimal $\mathbf{X}_w$ that minimizes the objective function, performed as an inner loop where we examine all possible choices of $\mathbf{X}_w$. Due to the specific structure of $\mathbf{X}_w$, this process is equivalent to solving the submatrix $\mathbf{x}_w$ within $\mathbf{X}_w$. In each iteration, two pieces of information are updated regarding $\mathbf{x}_w$, the position within $\mathbf{X}_w$ and the value of these elements.

Note that $\mathbf{D}$ represents a dictionary matrix that governs all elements in both $\mathbf{X}$ and $\mathbf{I}$ to be the same, with the exception of elements at the positions of $(i,i)$,$(i,j)$,$(j,i)$, and $(j,j)$. We can choose the position of $\mathbf{x}_w$ based on the choice of $\mathbf{D}$. Through adjusting the $\mathbf{D}$ in the equality constraint, $\mathbf{D} \mathbf{X} = \mathbf{D} \mathbf{I}$, we can determine the choice of the four elements within $\mathbf{X}_w$ that do not correspond to an identity matrix, thus the position of $\mathbf{x}_w$.

To determine the optimal $\mathbf{X}_w$, we employ an exhaustive search method, considering all feasible combinations of $(i,i)$, $(i,j)$, $(j,i)$, and $(j,j)$ by varying $\mathbf{D}$. We calculate the best value of $\mathbf{x}_w$ to minimize Eq.~\eqref{eq:lagrangian} for each potential position. Through a comprehensive assessment of all feasible solutions of $\mathbf{X}_w$, iterating over the positions of $\mathbf{x}_w$, we select the global optimal $\mathbf{X}_w$ and determine the position of $\mathbf{x}_w$ based on the $\mathbf{D}$ that yields the best solution.

Furthermore, since $\mathbf{D}$ and $I$ are known, we can represent their multiplication term as $C$, rewriting Eq.~\eqref{eq:lagrangian} as:
 \begin{equation} \label{eq:subprob}
\mathrm{arg \min_{\mathbf{X}_w}} \;\; \frac{1}{2}\left\lVert\mathbf{A} \mathbf{X}_w \mathbf{B} - \mathbf{U}\right\rVert^2_2 + \lambda (\mathbf{X}_w^T \mathbf{X}_w - \mathbf{I} ) \;\; \mathrm{s.t.} \;\; \mathbf{D} \mathbf{X}_w = 
\mathbf{C} \;\; .
\end{equation}

Note that by vectorization, the objective function Eq.~\eqref{eq:subprob} can be formulated as a least squares objective:
\begin{equation} \label{eq:lsobj}
\mathrm{arg \min_{\mathbf{X}_w}} \;\; \frac{1}{2}\left\lVert\mathrm{vec}(\mathbf{A} \mathbf{X}_w \mathbf{B} - \mathbf{U})\right\rVert^2_2 + \lambda (\mathrm{vec}(\mathbf{X}_w)^T \mathrm{vec}(\mathbf{X}_w) - \mathbf{I} )\;\;\;\; \mathrm{s.t.}\;\;  \mathbf{D}\mathbf{X}_w = \mathbf{C} \;\; ,
\end{equation}
which is equivalent to:
 \begin{equation} \label{eq:lsobj2}
\mathrm{arg \min_{\mathbf{X}_w}} \;\; \frac{1}{2}\left\lVert(\mathbf{A} \otimes \mathbf{B}^T) \mathrm{vec}(\mathbf{X}_w) - \mathrm{vec}(\mathbf{U})\right\rVert^2_2 + \lambda (\mathrm{vec}(\mathbf{X}_w)^T \mathrm{vec}(\mathbf{X}_w) - \mathbf{I} )\;\;\;\; \mathrm{s.t.}\;\;  \mathbf{D}\mathbf{X}_w = \mathbf{C} \;\;. 
\end{equation}

Consider $\mathbf{Z} =  \mathrm{vec}(\mathbf{X}_k$), Eq.~\eqref{eq:lsobj2} is equivalent to:
\begin{equation} \label{eq:lsobj3}
\mathrm{arg \min_{\mathbf{Z}} }\;\; \frac{1}{2} ((\mathbf{A} \otimes \mathbf{B}^T) \mathbf{Z} - \mathrm{vec}(\mathbf{U})) ((\mathbf{A} \otimes \mathbf{B}^T) \mathbf{Z} - \mathrm{vec}(\mathbf{U})) +\lambda (\mathbf{Z}^T \mathbf{Z} - \mathbf{I})\;\;\;\; \mathrm{s.t.}\;\;  \mathbf{D}\mathbf{Z} = \mathbf{C} \;\;.
\end{equation}
This can be further expanded to:
\begin{equation} \label{eq:flsobj}
\begin{split}
\mathrm{arg \min_{\mathbf{Z}}} \;\; \frac{1}{2} \mathbf{Z}^T ( (\mathbf{A} \otimes \mathbf{B}^T)^T(\mathbf{A} \otimes \mathbf{B}^T)) \mathbf{Z} +\lambda (\mathbf{Z}^T \mathbf{Z} - \mathbf{I})
\\- \mathrm{vec}(\mathbf{U})^T (\mathbf{A} \otimes \mathbf{B}^T) \mathbf{Z}  + \frac{1}{2} \mathrm{vec}(\mathbf{U})^T \mathrm{vec}(\mathbf{U}) \\
\mathrm{s.t.}\;\;  \mathbf{D}\mathbf{Z}= \mathbf{C} \;\; . 
\end{split}
\end{equation}

By considering, 
$\mathbf{Q} = (\mathbf{A} \otimes \mathbf{B}^T)^T(\mathbf{A} \otimes \mathbf{B}^T)$,
$\mathbf{p} = \mathrm{vec}(\mathbf{U})^T (\mathbf{A} \otimes \mathbf{B}^T)$
$\text{constant} = \frac{1}{2} \mathrm{vec}(\mathbf{U})^T \mathrm{vec}(\mathbf{U})$. Eq.~\eqref{eq:flsobj} can be considered as a constrained quadratic equation:

\begin{equation} \label{sub_final}
\begin{split}
\mathrm{arg \min_{\mathbf{Z}} } \;\; \frac{1}{2} \mathbf{Z}^{T} \mathbf{Q} \mathbf{Z} +\lambda (\mathbf{Z}^T \mathbf{Z} - \mathbf{I}) -\mathbf{p}^{T} \mathbf{Z}  
\\
\mathbf{D}\mathbf{Z} =  \mathbf{C} \;\;.
\end{split}
\end{equation}

For each iteration, we update the gate matrix $\mathbf{X}_{w} =  \mathbf{Z}^*$. 
We use the Karush-Kuhn-Tucker (KKT) conditions~\cite{kuhn2013nonlinear} to derive a system of linear equations. The KKT conditions are necessary conditions for a solution in nonlinear programming to be optimal, combining the objective function and constraints into a single system.
The solution, $\mathbf{Z}^*$, of quadratic problem Eq.~\eqref{sub_final} can solved as follows:

Considering the Lagrangian form of Eq.~\eqref{sub_final}
\begin{equation} \label{sub_l}
L(\mathbf{Z}) = \frac{1}{2}\mathbf{Z}^{T} \mathbf{Q} \mathbf{Z} +\lambda (\mathbf{Z}^T \mathbf{Z}- \mathbf{I})   - \mathbf{p}^{T} \mathbf{Z} +\mu (\mathbf{D}\mathbf{Z} - \mathbf{C} ) \;\;,
\end{equation}
with its derivative as,
\begin{equation} \label{sub_dl}
\nabla L(\mathbf{Z}) = \mathbf{Q} \mathbf{Z} + 2\lambda\mathbf{Z}  - \mathbf{p} + \mu \mathbf{D} = 0 \;\;.
\end{equation}

The quadratic problem is given as, 
\begin{equation} \label{sub_matrix}
 \begin{bmatrix} \mathbf{Q} + 2 \lambda \mathbf{I} & \mathbf{D}^{T} \\ \mathbf{D} & \mathbf{0} \end{bmatrix}
 \begin{bmatrix} \mathbf{Z} \\ \mu \end{bmatrix}
 =
  \begin{bmatrix}  \mathbf{p} \\ \mathbf{C} \end{bmatrix} \;\;.
\end{equation}

 Eq.~\eqref{sub_matrix} compose of two equations, the first row gives  Eq.~\eqref{sub_dl} and the second row is the constraint given in  Eq.~\eqref{sub_final}. 
 Obtaining the solution of $\mathbf{Z}^*$ of Eq.~\eqref{sub_matrix} leads to

\begin{equation} \label{z_sol}
\mathbf{Z}^* = (\mathbf{Q + 2\lambda \mathbf{I}})^{-1}(\mathbf{p} - \mathbf{D} \mu) \;\;.
\end{equation}

Furthermore, for the generalizability of our algorithm, we can perform calculations without imposing the unitary constraint. By removing the unitary constraint in Eq.~\eqref{eq:subprob}, Eq.~\eqref{sub_matrix} transforms into:
\begin{equation} \label{sub_matrix2}
 \begin{bmatrix} \mathbf{Q} & \mathbf{D}^{T} \\ \mathbf{D} & \mathbf{0} \end{bmatrix}
 \begin{bmatrix} \mathbf{Z} \\ \mu \end{bmatrix}
 =
  \begin{bmatrix}  \mathbf{p} \\ \mathbf{C} \end{bmatrix} \;\;, 
\end{equation}
with  the solution of $\mathbf{Z}^*$ as:
\begin{equation} \label{z_sol2}
\mathbf{Z}^* = (\mathbf{Q})^{-1}(\mathbf{p} - \mathbf{D} \mu) \;\;.
\end{equation}

\subsection{Update of penalty parameters}

Finally, we update the penalty parameters $\lambda$ and $\mu$ by adaptively based on the gradient of the function. Namely, in each iteration, considering the gradient norm $\| \nabla f(x_k) \|$:

If the gradient norm $\| \nabla f(x_k) \|$ is large, $\lambda_{k+1}$ and $\mu_{k+1}$ is decreased by $s_1$:
\begin{equation} \label{updatelm}
\begin{split}
    \lambda_{k+1} = s_{1} \lambda_k \\
    \mu_{k+1} = s_{1} \mu_k    
\end{split}
\end{equation}

Whereas, if the gradient norm $\| \nabla f(x_k) \|$ is small, $\lambda_{k+1}$ and $\mu_{k+1}$ is increased by $s_2$:
\begin{equation} \label{updatelm2}
\begin{split}
    \lambda_{k+1} = s_{2} \lambda_k \\
    \mu_{k+1} = s_{2} \mu_k    
\end{split}
\end{equation}

where $0< s_1 < 1$ in \eqref{updatelm} and $1< s_2<2$ in \eqref{updatelm2} are small increments and decreases for the penalty parameters. By adjusting $\lambda$ based on the magnitude of the gradient norm at each iteration, the algorithm dynamically tunes the penalty parameters to ensure efficient convergence. Decreasing $\lambda$ and $\mu$ when the gradient norm is large allows for more significant updates in such cases, while increasing $\lambda$ and $\mu$ when the gradient norm is small helps fine-tune the optimization process for smoother convergence towards the optimal solution.

\section{Numerical Simulation} \label{sec:result}
In this section, we present numerical simulations conducted on our proposed block decomposition approach for gate reduction to approximate a given transformation operator. The purpose of these simulations is to demonstrate the performance of our algorithm.

We randomly generated a collection of complex matrices in Matlab to represent target transformation matrices $\mathbf{U}$, which denote specific transformation operations we aim to apply to quantum states. The objective of this simulation is to test how well the unitary matrix $\mathbf{Y}$ that can approximate the provided $\mathbf{U}$.
All numerical simulations are implemented in MATLAB R2021a and executed on a personal computer with an Intel(R) Core(TM) i7-8750H CPU clocked at 2.21 GHz and 16 GB RAM.

\subsection{Approximation performance}

\begin{table} [h]
\centering
    \begin{tabular}{|c|c|c|c|c|c|}
    \hline
    \textbf{Number of Gates} &  M=$5$ & M=$10 $& M=$15$ & M= $20 $& M=$25$  \\
    \hline
    {Converged value of $L$} & $4.51$ & $3.87$ & $3.21 $& $2.31$ & $1.83$ \\
    \hline
    {No. of iterations required for convergence} & $68$ & $110$ & $151$ & $137$ & $128$ \\
    \hline
    {Time required for convergence  (seconds)} & $5.05$ & $8.16 $&$ 16.01$ & $12.08$ &$ 12.59$  \\
    \hline
    \end{tabular}
     
\caption{Algorithm performance across varying numbers of gates for approximation on 3 qubits circuit.}
\label{tab:Gate_test}
\end{table}

\begin{table} [h]
\centering
    \begin{tabular}{|c|c|c|c|}
    \hline
    \textbf{Number of Qubits} &  $3$& $4$& $5$  \\
    \hline
    {Time required for each iteration (minutes) } & $<1$ & $<15$ & $>30$ \\
    \hline
    \end{tabular}
\caption{Algorithm performance across different number of qubits, with $M=10$.}
\label{tab:size_test}
\end{table}

This subsection aims to showcase the performance of the proposed algorithm by selecting different numbers of gates to approximate a given transformation matrix. Numerous simulations were conducted to ensure result consistency and algorithm convergence.  The study focuses on a 3-qubit system, where any transformation can be depicted by a circuit with a maximum of $28$ gates~\cite{li2014decomposition}, with each of these gates capturing distinct facets of the transformation process. In this simulation, we construct random target unitary matrices $\mathbf{U}$ by multiplying $28$ unitary $\mathbf{U}_k$ gates randomly generated in Matlab and approximate $\mathbf{U}$ with transformation matrices $\mathbf{Y}$ containing only $M$ gates through our algorithm. Table~\ref{tab:Gate_test} compares the behavior and effectiveness of the approximation with varying values of $M$. Each data point in the tables represents an average of 30 trials.

The table shows the performance across varying numbers of gates for approximating a 3-qubit transformation. Increasing the gate count results in a more accurate approximation of the target unitary matrix, as reflected in the decreasing converged values of $L$. However, this heightened precision comes at the cost of increased computational complexity. With a larger number of gates, the algorithm shows a corresponding increase in the number of iterations and convergence time. Nonetheless, these outcomes underscore the algorithm effective convergence within a constrained timeframe and iteration count, showcasing its efficiency in navigating complex transformations with precision.
These results highlight the trade-offs and considerations involved in selecting the appropriate gate count for approximating the transformation matrix. 

Furthermore, Table~\ref{tab:size_test} displays a limitation of our algorithm. The table illustrates the average time required to update each $\mathbf{X}_k$ within a single iteration. Given that the number of possible positions for exhaustive search remains constant for matrices of the same size, we can extrapolate the optimization completion time for varying numbers of qubits by multiplying the time per iteration by the total number of iterations needed for convergence. 

From the table, we see that as the number of qubits increases, the time to converge becomes impractical. This is because increasing the number of qubits by 1 doubles the size of the transformation matrix, resulting in a matrix with dimensions of $2^{(n+1)} \times 2^{(n+1)}$. The increased size of the transformation matrix introduces a wide variety of sub-matrix positions, leading to a significant increase in the number of quadratic programming sub-problems required, thereby prolonging the computational time. This observation underscores the challenges associated with scaling up to larger quantum systems. 

A promising solution to address this bottleneck involves the introduction of indicator variables for dictionary selection. This strategic implementation enables the transformation of quadratic programming sub-problems into binary quadratic problems with linear constraints. This modification significantly reduces the computational complexity, leading to faster processing times and more efficient computation.

\subsection{Circuit Formulation}

\begin{figure} 
  \centering
\includegraphics[width=0.68\columnwidth]{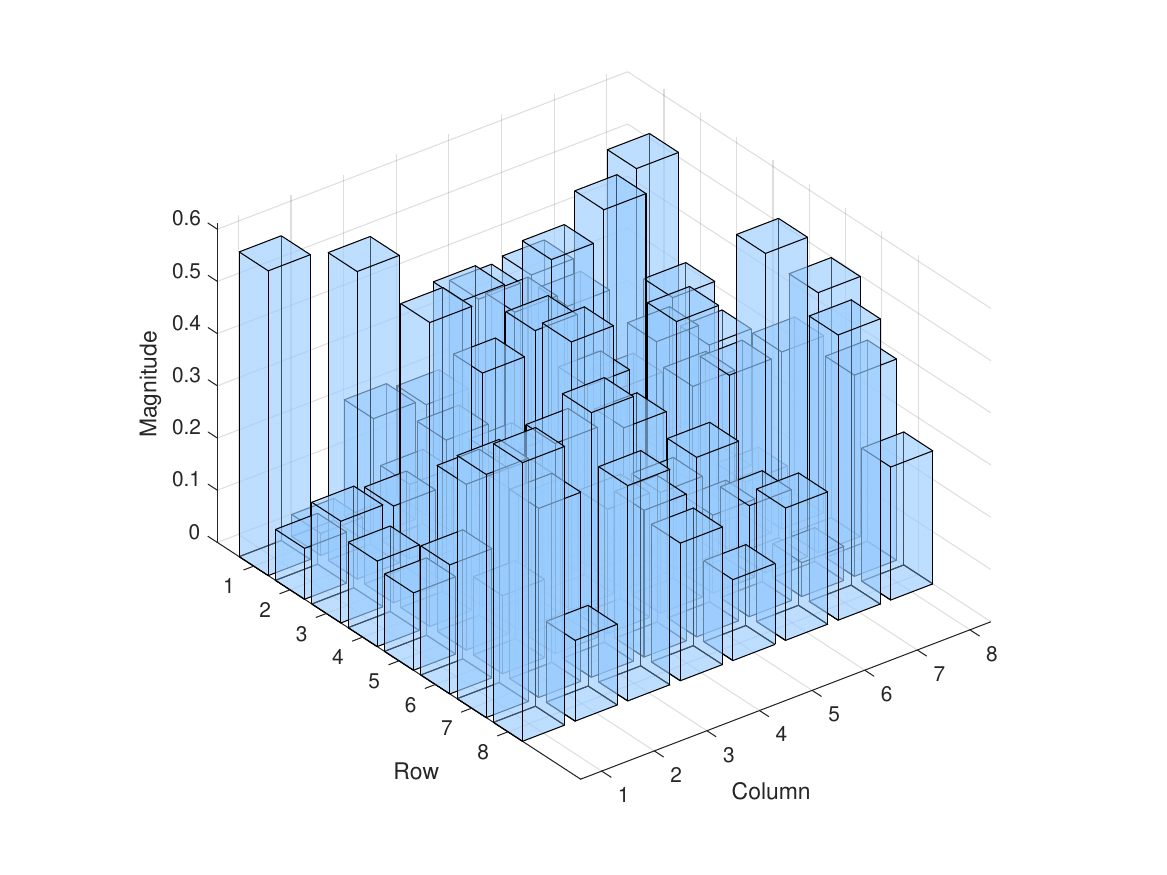}
  \caption{Magnitude of example unitary transformation matrix, $\mathbf{U}$ in Eq.~\eqref{eq:example}.}
  \label{fig:example}
\end{figure}

\begin{figure} 
  \centering
\includegraphics[width=0.9\columnwidth]{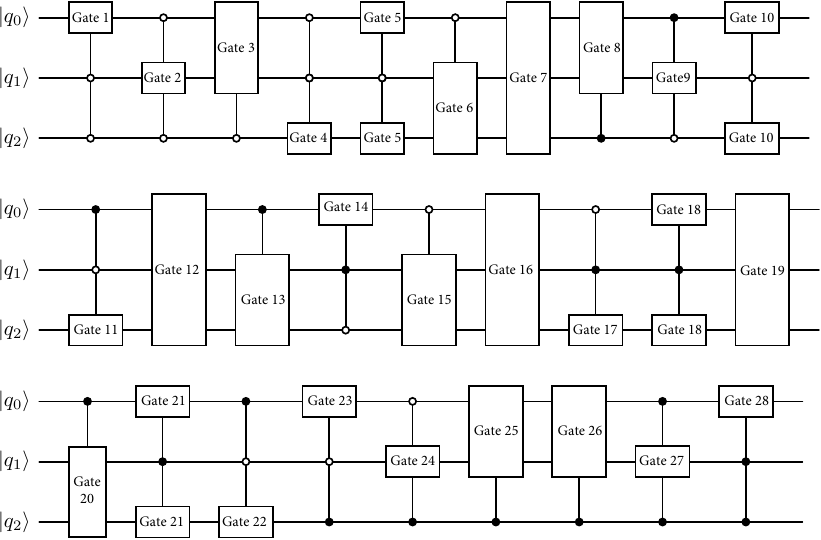}
  \caption{General form of quantum circuit to represent $3$-qubit transformation matrix $\mathbf{U}$ in Eq.~\eqref{eq:example}.}
  \label{fig:circuit}
\end{figure}

In this section, we present an illustration involving a 3-qubit system, showcasing the process of approximating a random transformation matrix and constructing the corresponding quantum circuit. Beginning with a randomly generated transformation matrix in MATLAB, denoted as matrix $\mathbf{U}$ given in Eq.~\eqref{eq:example}, with detail magnitude visually shown in Fig.~\ref{fig:example}, we translate this matrix into a quantum circuit depicted in Fig.~\ref{fig:circuit}. Each gate in this circuit is composed of a sequence of $R_z$, $R_y$, and $R_z$ gates, with the specific rotational angles detailed in Table~\ref{Tab:28gate}.
This example serves as a concrete illustration of the algorithm's capability in reducing quantum gates in real-world applications.

{\color{black}
\begin{equation} \label{eq:real}
U_{\text{real}} = \begin{bmatrix}
-0.113 & -0.041 & 0.223 & -0.192 & -0.308 & -0.321 & 0.051 & 0.509 \\
-0.044 & -0.556 & -0.142 & 0.216 & -0.256 & 0.310 & 0.103 & 0.027 \\
-0.057 & 0.183 & -0.031 & -0.180 & 0.059 & -0.423 & -0.199 & 0.062 \\
0.141 & -0.582 & -0.168 & -0.455 & 0.151 & 0.020 & -0.256 & -0.035 \\
0.102 & -0.074 & 0.187 & 0.347 & 0.280 & -0.348 & 0.041 & -0.191 \\
0.196 & -0.014 & 0.378 & 0.048 & 0.021 & 0.139 & 0.040 & -0.178 \\
0.464 & 0.362 & -0.365 & 0.280 & -0.094 & 0.152 & 0.057 & 0.267 \\
0.319 & 0.022 & 0.334 & 0.056 & 0.154 & -0.025 & -0.209 & -0.255
\end{bmatrix}
\end{equation}

\begin{equation} \label{eq:imaginary}
U_{\text{imaginary}} = \begin{bmatrix}
-0.573 & 0.008 & -0.006 & 0.089 & 0.214 & -0.208 & 0.122 & -0.011 \\
-0.087 & -0.195 & -0.002 & 0.401 & -0.301 & -0.197 & -0.138 & -0.306 \\
0.186 & 0.029 & -0.271 & -0.088 & -0.538 & -0.423 & -0.234 & -0.254 \\
0.083 & 0.020 & -0.414 & -0.181 & 0.310 & -0.058 & 0.073 & 0.026 \\
0.108 & -0.309 & -0.179 & 0.378 & 0.132 & -0.323 & 0.331 & 0.280 \\
-0.151 & 0.149 & -0.139 & -0.233 & -0.230 & 0.055 & 0.610 & -0.465 \\
0.038 & 0.001 & -0.352 & 0.021 & 0.331 & -0.149 & -0.031 & -0.278 \\
-0.428 & 0.153 & -0.242 & 0.258 & -0.018 & 0.252 & -0.505 & -0.004
\end{bmatrix}
\end{equation}

\begin{equation} \label{eq:example}
U = U_{\text{real}}  \;\;+ \;\; U_{\text{imaginary}}
\end{equation}
}

\begin{table}[H]
\centering
{ \color{black}
\begin{tabular}{|c|c|c|c|}
\hline
Gate & $\theta$& $\phi$ & $\lambda$ \\
\hline
Gate 1 & $0.166$ & $0.166$ & $0.663$ \\
\hline
Gate 2 & $0.318$ & $0.318$ & $1.274$\\
\hline
Gate 3 & $0.257$ & $0.257$ & $2.611$\\
\hline
Gate 4 & $0.226$ & $0.226$ & $2.326$\\
\hline
Gate 5 & $0.358$ & $0.358$ & $-2.487$\\
\hline
Gate 6 & $0.583$ & $0.583$ & $3.061$ \\
\hline
Gate 7 & $0.563$ & $0.563$ & $-2.212$\\
\hline
Gate 8 & $0.251$ & $0.251$ & $1.443$\\
\hline
Gate 9 & $0.788$ & $0.788$ & $-3.063$\\
\hline
Gate 10 & $0.329$ & $0.329$ & $1.832$ \\
\hline
Gate 11 & $0.167$ & $0.167$ & $-1.258$\\
\hline
Gate 12 & $0.405$ & $0.405$ & $-0.111$ \\
\hline
Gate 13 & $0.184$ & $0.184$ & $-1.425$\\
\hline
Gate 14 & $1.098$ & $1.098$ & $1.411$\\
\hline
Gate 15 & $0.389$ & $0.389$ & $-2.669$\\
\hline
Gate 16 & $0.776$ & $0.776$ & $-2.629$ \\
\hline
Gate 17 & $0.656$ & $0.656$ & $-0.932$\\
\hline
Gate 18 & $0.519$ & $0.519$ & $-2.515$\\
\hline
Gate 19 & $1.328$ & $1.328$ & $1.415$\\
\hline
Gate 20 & $0.244$ & $0.244$ & $-1.239$\\
\hline
Gate 21 & $0.608$ & $0.608$ & $-0.209$\\
\hline
Gate 22 & $0.374$ & $0.374$ & $-1.358$\\
\hline
Gate 23 & $0.469$ & $0.469$ & $-0.229$\\
\hline
Gate 24 & $0.64$ & $0.64$ & $1.374$\\
\hline
Gate 25 & $0.233$ & $0.233$ & $-3.026$ \\
\hline
Gate 26 & $0.592$ & $0.592$ & $-3.049$\\
\hline
Gate 27 & $0.398$ & $0.398$ & $-1.668$\\
\hline
Gate 28 & $1.134$ & $1.134$ & $-1.926$\\
\hline
\end{tabular}

}
\caption{Gate angles of quantum circuit for constructing transformation matrix $\mathbf{U}$ in Eq.~\eqref{eq:example}.}
\label{Tab:28gate}
\end{table}

\begin{table} [H]
\centering

Example 1:
{\color{black}
\begin{tabular}{|c|c|c|c|c|c|}
\hline
Position $i$ & Position $j$ & Corresponding Gate &  $\theta$& $\phi$ & $\lambda$ \\
\hline
$4$ & $6$ & Gate 20 & $0.641$ & $0.326$ & $-0.774$ \\
\hline
$2$ & $6$ & Gate 11 & $0.519$ & $0.134$ & $1.362$\\
\hline
$4$ & $5$ & Gate 19 & $0.339$ & $0.955$ & $2.071$\\
\hline
$2$ & $7$ & Gate 12 & $0.586$ & $0.752$ & $-2.34$\\
\hline
$5$ & $8$ & Gate 25 & $0.93$ & $0.987$ & $3.02$\\
\hline
$3$ & $4$ & Gate 14 & $0.47$ & $0.542$ & $1.661$\\
\hline
$6$ & $7$ & Gate 26 & $1.034$ & $1.048$ & $0.384$\\
\hline
$3$ & $5$ & Gate 15 & $1.377$ & $0.354$ & $1.704$\\
\hline
$5$ & $7$ & Gate 24 & $1.433$ & $0.904$ & $-2.842$\\
\hline
$1$ & $2$ & Gate 1 & $0.785$ & $0.785$ & $1.088$\\
\hline
\end{tabular} }

\vspace{20pt}

Example 2:
{\color{black}
\begin{tabular}{|c|c|c|c|c|c|}
\hline
Position $i$ & Position $j$ & Corresponding Gate & $\theta$& $\phi$ & $\lambda$ \\
\hline
$1$ & $6$ & Gate 5 & $0.047$ & $1.433$ & $1.227$ \\
\hline
$1$ & $4$ & Gate 3 & $0.78$ & $0.476$ & $-1.333$\\
\hline
$2$ & $4$ & Gate 9 & $0.257$ & $0.257$ & $2.611$\\
\hline
$2$ & $7$ & Gate 12 & $1.199$ & $0.649$ & $0.894$\\
\hline
$2$ & $5$ & Gate 10 & $0.329$ & $0.329$ & $1.832$\\
\hline
$2$ & $8$ & Gate 13 & $0.167$ & $0.812$ & $1.062$\\
\hline
$4$ & $5$ & Gate 19 & $0.973$ & $0.954$ & $-2.208$\\
\hline
$5$ & $6$ & Gate 23 & $1.328$ & $1.328$ & $1.415$\\
\hline
$2$ & $6$ & Gate 11 & $1.136$ & $1.168$ & $0.377$\\
\hline
$3$ & $7$ & Gate 17 &$1.186$ & $1.362$ & $2.529$\\
\hline

\end{tabular}}
\caption{Examples of gate angles in quantum circuit for constructing transformation matrix 
$\mathbf{U}$ in Eq.~\eqref{eq:example} using only $10$ gates}
\label{Tab:10gate}
\end{table}


{\color{black} Table~\ref{Tab:28gate} and Table~\ref{Tab:10gate} explicitly display the angles $\theta$, $\phi$, and $\lambda$ corresponding to the gates in Eq.~\eqref{eq:rotation}, with matrix forms as defined in Eq.~\eqref{eq:rotation_Rzy}.} Following the sequence and angles of the gates outlined in Table~\ref{Tab:10gate}, we can derive an approximation circuit $\mathbf{Y}$ of the circuit corresponding to matrix in Eq.~\eqref{eq:example}, with details of the corresponding complete circuit shown in Table~\ref{Tab:28gate}. Each individual gate is visually represented in Fig.~\ref{fig:circuit}, illustrating the structure of the gates used to build a circuit for a 3-qubit system.  Our algorithm considers all possible positions for $\mathbf{X}_w$ through a combinatorial search across all the gates shown in Fig.~\ref{fig:circuit}. Consequently, the algorithm will require more computing time for a higher number of qubits, since more gates need to be searched.

Note that Table~\ref{Tab:10gate} presents two examples of the converged result for our matrix approximation in Eq.~\eqref{eq:example}, minimizing the number of gates of the circuit shown in Table~\ref{Tab:28gate}. 
Rather than having a straightforward single global minimum, this problem is characterized by multiple local optima. However, our algorithm consistently reaches an optimal point for the loss function in Eq.~\eqref{eq:main}. This was validated through a series of $30$ trials, ensuring robustness and reliability in the results obtained. The algorithm repeatedly converged with the loss function pinpointed at $3.773$ for the transformation matrix in Eq.~\eqref{eq:example}, showcasing the consistency and precision of our approach.

The behavior of our algorithm is directly influenced by the initial selection of the optimization gates, as detailed in Table~\ref{Tab:10gate}. Although both examples reach the same optimal loss function, they do so through different gate orders and values. Notably, certain gates, such as Gates 10, 13, and 18 in example 1, and Gates 11, 13, 14, 15, and 18 in example 2, remain unchanged. This phenomenon arises from their placement in the later stages of the block decomposition process. As prior gates have already been updated based on their values, these specific gates retain their original rotation settings.

While the block decomposition method can achieve a global minimum for $\ell_0$ sparse optimization~\cite{yuan2020block}, this is due to the non-convex nature of the objective arising solely from the $\ell_0$ constraint. This method decomposes the non-convex $\ell_0$ constraint into chunks and searches for the global solution within each chunk. In contrast, our problem exhibits dual complexities contributing to the non-convexity of the function. While the positioning of sub-matrices is globally explored using a similar methodology~\cite{yuan2020block}, the unitary constraint introduces additional layers of non-convexity, leading to the existence of multiple local optima.

Moreover, the approximated sub-matrices that represent each gate are designed to be nearly unitary matrices. The unitarity of these gates is significantly influenced by the penalty parameter $\lambda$ in Eq.~\eqref{eq:lagrangian}. Opting for a small penalty parameter may result in the matrix failing to uphold the unitary constraint. Conversely, selecting a high penalty value can force the resultant matrix to become an identity matrix due to excessive restrictions.

\begin{figure} 
  \centering
\includegraphics[width=0.68\columnwidth]{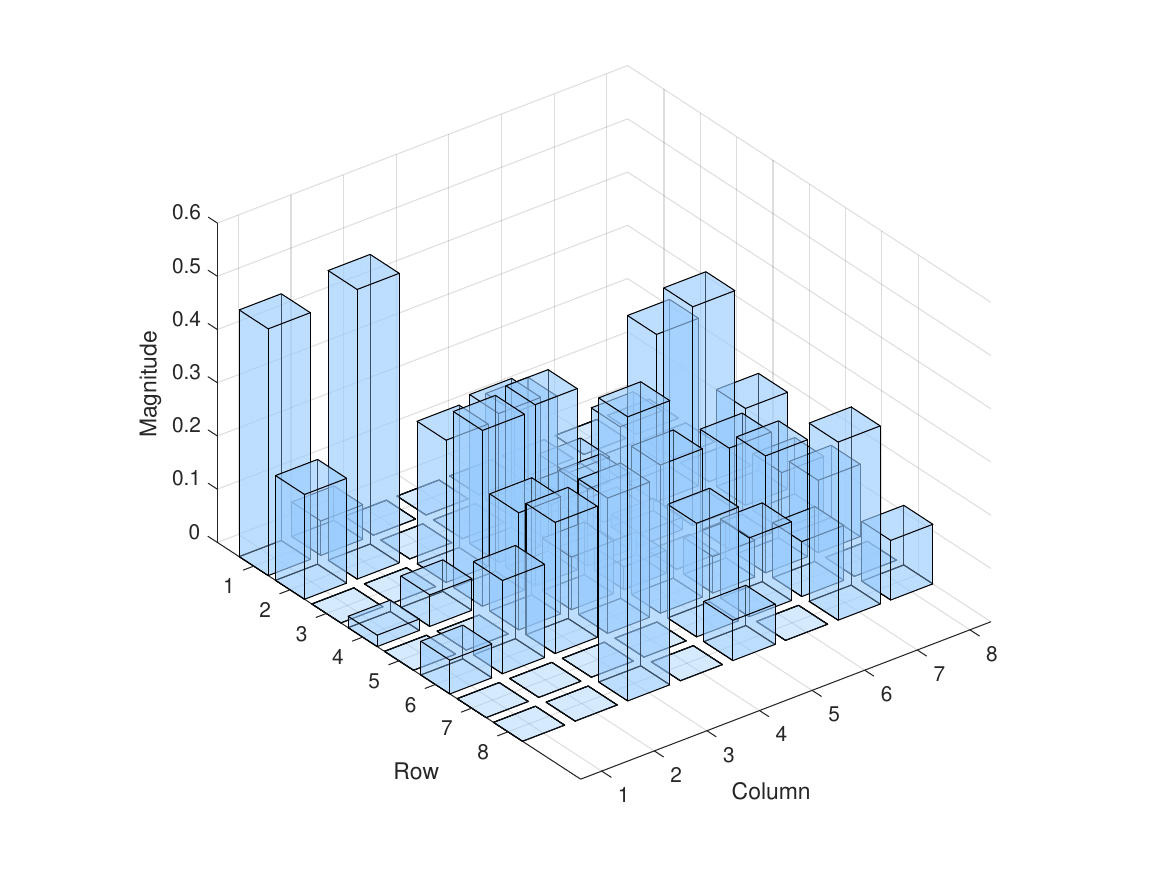}
\caption{Magnitude of resulted transformation matrix built using $10$ gates, $\mathbf{Y}$.}
  \label{fig:10gate}
\end{figure}

By combining the $10$ gates in Example 1 of Table~\ref{Tab:10gate}, a quantum circuit is formed to represent the transformation matrix Eq.~\eqref{eq:example}. Comparing the magnitudes of the original matrix in Fig.~\ref{fig:example}and the resulting matrix in Fig.~\ref{fig:10gate} obtained using the proposed algorithm, it is evident that the transformation matrix is sparse. Comparing this sparseness to matrices obtained through the introduction of sparsity penalties such as $\ell_1$ and $\ell_{2,1}$ norms~\cite{man2024group} this method drastically reduces the number of gates needed to approximate a given transformation matrix.

Next, we evaluate the performance of the approximated transformation matrix. The transformation matrix $\mathbf{U}$ aims to convert an initial state $|\psi_{i}\rangle$ to a target state $|\psi_{t}\rangle$, such that $\mathbf{U} |\psi_{i}\rangle = |\psi_{t}\rangle$. Taking Example 1 from~\ref{Tab:10gate}, our goal is to test whether the approximated transformation matrix $\mathbf{Y}$ can effectively replace $\mathbf{U}$ in this process, such that $\mathbf{Y} |\psi_{i}\rangle \approx |\psi_{t}\rangle$.

Consider the W state given by
\begin{equation} \label{eq:Wstate}
    |W\rangle = \frac{1}{\sqrt{3}} (|001\rangle + |010\rangle + |100\rangle)\;\; .
\end{equation}

We consider $\mathbf{U}$ on state Eq.~\eqref{eq:Wstate}, $U|W\rangle$ is 
 {\color{black}
\begin{equation} \label{eq:UWstate}
\begin{split}
\mathbf{U} |W\rangle = &\left(-0.019 + 0.0191i\right) |000\rangle + \left(-0.171 - 0.254i\right) |001\rangle \\
&+ \left(-0.099 - 0.602i\right) |010\rangle + \left(-0.058 - 0.018i\right) |011\rangle \\
&+ \left(0.293 + 0.163i\right) |100\rangle + \left(0.253 + 0.139i\right) |101\rangle \\
&+ \left(-0.232 - 0.031i\right) |110\rangle + \left(0.161 + 0.441i\right) |111\rangle
\end{split}
\end{equation}
}

For instance, randomly selecting a subset of 10 gates from $\mathbf{U}$ results in $\mathbf{Y}_0$, which provides a poor estimation of state in Eq.~\eqref{eq:UWstate}, given as:
 {\color{black}
\begin{equation} \label{eq:V0Wstate}
\begin{split}
\mathbf{Y}_0 |W\rangle = & \left(-0.012 + 0.026i\right) |000\rangle + \left(-0.020 + 0.090i\right) |001\rangle \\
&+ \left(-0.528 + 0.005i\right) |010\rangle + \left(0.488 - 0.188i\right) |011\rangle \\
&+ \left(-0.043 - 0.100i\right) |100\rangle + \left(-0.136 + 0.345i\right) |101\rangle \\
&+ \left(0.454 + 0.153i\right) |110\rangle + \left(-0.196 + 0.142i\right) |111\rangle
\end{split}
\end{equation}}
    
Optimizing $\mathbf{Y}_0$, the resultant matrix $\mathbf{Y}$, constructed with $10$ gates, predicts the state in Eq.~\eqref{eq:UWstate} as:
 {\color{black}
\begin{equation} \label{eq:VWstate}
\begin{split}
\mathbf{Y} |W\rangle = &\left(0\right) |000\rangle + \left(-0.036 - 0.089i\right) |001\rangle \\
&+ \left(0.015 - 0.109i\right) |010\rangle + \left(-0.058 - 0.018i\right) |011\rangle \\
&+ \left(-0.178 + 0.125i\right) |100\rangle + \left(0.349 + 0.222i\right) |101\rangle \\
&+ \left(-0.125 - 0.099i\right) |110\rangle + \left(0.191 + 0.419i\right) |111\rangle
\end{split}
\end{equation}
}

{\color{black} To demonstrate the effectiveness of our approach, we evaluate the fidelity of both the resulting matrices $\mathbf{Y}_0$ and $\mathbf{Y}$ with respect to the target matrix $\mathbf{U}$.} 
The fidelity can be calculated by introducing the density matrices $|\psi\rangle = \mathbf{Y} |W\rangle$ and $|\phi\rangle = \mathbf{U} |W\rangle$, {\color{black} and then applying} the fidelity formula:

\begin{equation} 
F(|\psi\rangle, |\phi\rangle) = |\langle \psi | \phi \rangle|^2
\label{eq:fidelity}
\end{equation}  

{\color{black} Since our method can potentially be applied to any quantum operator, such as sparse unitary transformation matrices produced by the method in~\cite{man2024group}, applying these operators may lead to unnormalized states due to approximations. The comparison between the original transformation $\mathbf{U} |W\rangle$ and the approximations $\mathbf{Y}_0 |W\rangle$ and $\mathbf{Y}_0 |W\rangle$ showcases the impact of gate optimization on accuracy. To ensure meaningful comparisons, we first normalize the states $\mathbf{U} |W\rangle$~\eqref{eq:UWstate}, $\mathbf{Y}_0 |W\rangle$~\eqref{eq:V0Wstate}, and $\mathbf{Y}_0 |W\rangle$~\eqref{eq:VWstate} as follows: 

\begin{equation} \label{normalize}
|\psi_{\text{normalized}}\rangle = \frac{1}{\sqrt{\sum_{n} |c_n|^2}} \sum_{n} c_n |n\rangle.
\end{equation}

By applying~\eqref{normalize} to $\mathbf{U} |W\rangle$~\eqref{eq:UWstate}, $\mathbf{Y}_0 |W\rangle$~\eqref{eq:V0Wstate}, and $\mathbf{Y}_0 |W\rangle$~\eqref{eq:VWstate}, the normalized states are given as:

\begin{equation} \label{UWnorm}
\begin{split}
U|W\rangle_{\text{normalized}} = &\frac{1}{0.9702} \big[ (-0.019 + 0.0191i) |000\rangle + (-0.171 - 0.254i) |001\rangle \\
&+ (-0.099 - 0.602i) |010\rangle + (-0.058 - 0.018i) |011\rangle \\
&+ (0.293 + 0.163i) |100\rangle + (0.253 + 0.139i) |101\rangle \\
&+ (-0.232 - 0.031i) |110\rangle + (0.161 + 0.441i) |111\rangle \big],
\end{split}
\end{equation}

\begin{equation} \label{V0Wnorm}
\begin{split}
Y_0|W\rangle_{\text{normalized}} = &\frac{1}{0.9995} \big[ (-0.012 + 0.026i) |000\rangle + (-0.020 + 0.090i) |001\rangle \\
&+ (-0.528 + 0.005i) |010\rangle + (0.488 - 0.188i) |011\rangle \\
&+ (-0.043 - 0.100i) |100\rangle + (-0.136 + 0.345i) |101\rangle \\
&+ (0.454 + 0.153i) |110\rangle + (-0.196 + 0.142i) |111\rangle \big],
\end{split}
\end{equation}

\begin{equation} \label{VWnorm}
\begin{split}
Y|W\rangle_{\text{normalized}} = &\frac{1}{0.6935} \big[ (0) |000\rangle + (-0.036 - 0.089i) |001\rangle \\
&+ (0.015 - 0.109i) |010\rangle + (-0.058 - 0.018i) |011\rangle \\
&+ (-0.178 + 0.125i) |100\rangle + (0.349 + 0.222i) |101\rangle \\
&+ (-0.125 - 0.099i) |110\rangle + (0.191 + 0.419i) |111\rangle \big].
\end{split}
\end{equation}

The fidelity between the normalized states is calculated as follows:

\begin{equation} \label{fidV0}
F(Y_0|W\rangle, U|W\rangle) = |\langle Y_0|W_{\text{normalized}} | U|W_{\text{normalized}} \rangle|^2 \approx 0.853.
\end{equation}

\begin{equation}  \label{fidV}
F(Y|W\rangle, U|W\rangle) = |\langle Y|W_{\text{normalized}} | U|W_{\text{normalized}} \rangle|^2 	\approx 0.921.
\end{equation}

By using Eq.~\eqref{eq:fidelity}, the fidelity of $\mathbf{Y}_0$ and $\mathbf{U}$ is $0.853$ from ~\eqref{fidV0}, whereas the fidelity between $\mathbf{Y}$ and $\mathbf{U}$ is $0.921$ from ~\eqref{fidV}.} {\color{black} Normalization ensures that the states represent valid quantum states, and the fidelity calculation reflects the true overlap between the states.} Since fidelity measures the closeness of two quantum states, higher fidelity indicates that the two states are more similar, this showed that $\mathbf{Y}$ yield a better result.

{\color{black} This comparison demonstrates that our optimization process significantly improves the approximation of the target transformation.} Extracting 10 gates randomly from $\mathbf{U}$ to form $\mathbf{Y}_0$ yields a notably poor estimate, deviating significantly from the true transformation $\mathbf{U}|W\rangle$. However, through optimization using the proposed algorithm, the refined transformation matrix $\mathbf{Y}$ constructed with the same number of gates is remarkably aligned closer to $\mathbf{U}|W\rangle$. This enhancement underscores the merit gains from optimization in the approximation of complex unitary transformations.

 \section{Discussion} \label{sec:discussion}
Our work builds upon and extends several existing methods in quantum circuit optimization and gate reduction. Traditional approaches to quantum circuit decomposition, such as those based on the Cosine-Sine Decomposition method~\cite{mottonen2004quantum, di2013synthesis}, focus on breaking down unitary matrices into sequences of elementary gates. Although these methods are effective in simplifying quantum circuits, they often result in a large number of gates, especially for high-dimensional transformations. In contrast, our approach introduces flexibility by allowing users to specify the desired number of gates, thereby optimizing gate usage while maintaining computational accuracy. This is particularly advantageous in scenarios where the gate count is a critical resource constraint.

In the field of optimization, recent work by \cite{yuan2020block} introduced the block decomposition algorithm, which has shown remarkable success in sparse optimization problems. Our method adapts this approach specifically for quantum gate reduction by incorporating unitary constraints and gate structure constraints. This adaptation allows us to achieve a balance between gate count reduction and transformation accuracy, which is not explicitly addressed in the original block decomposition framework. 
Furthermore, our work differs from gate reduction strategies that rely on automated optimization techniques, such as those discussed in~\cite{Nam_2018} and~\cite{abdessaied2014quantum}. Although these methods aim to reduce gate counts through iterative refinement and library-based optimization, they do not provide explicit control over the number of gates used in the final circuit. Our algorithm, on the other hand, allows users to directly specify the maximum number of gates, offering a more flexible and resource-efficient approach to quantum circuit design.

The significance of our work lies in its ability to bridge the gap between sparse optimization techniques and quantum circuit design. By integrating the block decomposition method with unitary constraints, we provide an alternative approach to approximating quantum transformations with a limited number of gates. This integration introduces flexibility in resource allocation, enabling users to tailor the gate count to specific hardware limitations or application requirements.

This flexibility represents a significant improvement over traditional decomposition methods, which typically produce circuits with a fixed number of gates based on the dimensionality of the transformation matrix. In summary, our work contributes to the growing body of research on quantum circuit optimization by introducing a new method that combines the strengths of block decomposition with the specific constraints of quantum gate design. This approach not only enhances the efficiency of quantum computations but also provides a more adaptable framework for designing quantum circuits in resource-constrained environments. 

\section{Conclusion} \label{sec:conclusion}
In conclusion, our paper presented an alternative method for approximating complex unitary transformations with a focus on limiting the number of gates. Considering the block decomposition technique, we successfully sparsed the transformation matrix with an adjustable gate count choice and achieved good approximation accuracy. Through simulations involving a 3-qubit transformation, we demonstrated the effectiveness of our approach and showcased the trade-offs between accuracy and computational complexity. Our study not only highlighted the manageable computational complexity but also underscored the limitations as the quantum system scales up in qubit numbers. 

\section*{Acknowledgments}

This work is supported by the National Natural Science Foundation of China (Grant No. 11874312, 12474489), Shenzhen Fundamental Research Program (Grant No. JCYJ20240813153139050), the Guangdong Provincial Quantum Science Strategic Initiative (Grant No. GDZX2203001, GDZX2403001), and the Innovation Program for Quantum Science and Technology (Grant No. 2021ZD0302300).

The data that support the findings of this article are openly available \cite{dataavail}.

\appendix
\section{Quantum Gate Positioning and Transformation Matrix Interpretation} \label{appx:appxA}

In quantum systems, the transformation matrix represents the evolution of quantum states with the transformation is applied to a given state, with each entry denoting the transition amplitude between specific quantum states. This data is pivotal for accurately mapping the gates onto the qubits in the quantum circuit.

The positioning of each $2 \times 2$ unitary matrix $\mathbf{u}_k$ within $\mathbf{U}_k$ determines the qubits on which the corresponding gate operates, elucidating the specific qubits that the transformation matrix $\mathbf{U_k}$ is applied.

When $\mathbf{u}_k$ is located at $(i,i)$, $(i, j)$, $(j, i)$, and $(j,j)$ positions, it indicates that the amplitudes for transitioning between the $i$-th and $j$-th input states correspond to the $i$-th and $j$-th output states, respectively. Ref.~\cite{li2014decomposition} provides a list of potential gate  for systems ranging from two to four qubits.
For instance, in a 3-qubit setup, the element at position $(i, j)$ in the unitary matrix signifies the transition amplitude from state $|j-1\rangle$ to state $|i-1\rangle$, contingent on the placement of $\mathbf{u}_k$. The transformation can be depicted through an $8 \times 8$ matrix as follows:

\begin{equation}  \label{eq:3x3matrix}
\begin{aligned}
\begin{array}{c|cccccccc}
    & |000\rangle & |001\rangle & |010\rangle & |011\rangle & |100\rangle & |101\rangle & |110\rangle & |111\rangle \\
\hline
|000\rangle & u_{1,1} & u_{1,2} & u_{1,3} & u_{1,4} & u_{1,5} & u_{1,6} & u_{1,7} & u_{1,8} \\
|001\rangle & u_{2,1} & u_{2,2} & u_{2,3} & u_{2,4} & u_{2,5} & u_{2,6} & u_{2,7} & u_{2,8} \\
|010\rangle & u_{3,1} & u_{3,2} & u_{3,3} & u_{3,4} & u_{3,5} & u_{3,6} & u_{3,7} & u_{3,8} \\
|011\rangle & u_{4,1} & u_{4,2} & u_{4,3} & u_{4,4} & u_{4,5} & u_{4,6} & u_{4,7} & u_{4,8} \\
|100\rangle & u_{5,1} & u_{5,2} & u_{5,3} & u_{5,4} & u_{5,5} & u_{5,6} & u_{5,7} & u_{5,8} \\
|101\rangle & u_{6,1} & u_{6,2} & u_{6,3} & u_{6,4} & u_{6,5} & u_{6,6} & u_{6,7} & u_{6,8} \\
|110\rangle & u_{7,1} & u_{7,2} & u_{7,3} & u_{7,4} & u_{7,5} & u_{7,6} & u_{7,7} & u_{7,8} \\
|111\rangle & u_{8,1} & u_{8,2} & u_{8,3} & u_{8,4} & u_{8,5} & u_{8,6} & u_{8,7} & u_{8,8} \\
\end{array}
\end{aligned}
\end{equation}

\end{document}